\documentclass[prapplied,aps,reprint,english,twocolumn,notitlepage]{revtex4-1}
\usepackage[T1]{fontenc}  \usepackage[latin1]{inputenc}
 \usepackage{babel} \usepackage{txfonts}
 \usepackage{epsfig,epsf,psfrag} 
\usepackage{graphicx} 
\usepackage{subfigure}
\usepackage{pslatex}
\usepackage{epic,eepic} 
\usepackage{color,pstcol}
\usepackage{pstricks} 
\usepackage{fancyhdr}  \parindent0em  
 \vspace{-10.0cm}  
\begin{document} \title{Fragility of non-local edge mode transport in the quantum spin Hall state} 
  \author{Arjun Mani}
 \author{Colin Benjamin} \email{colin.nano@gmail.com}\affiliation{National institute of Science education \& Research, Bhubaneswar 751005, India }
\begin{abstract}
 Non-local currents and voltages are better able at withstanding the deleterious effects of dephasing than local currents and voltages in nanoscale systems. This hypothesis is known to be true in quantum Hall set-ups. We test this hypothesis in a four terminal quantum spin Hall set up wherein we compare the local resistance measurement with the non-local one. In addition to inelastic scattering induced dephasing we also test resilience of the resistance measurements in the aforesaid set-ups to disorder and spin-flip scattering. We find the axiom that non-local resistance is less affected by the detrimental effects of disorder and dephasing to be in general untrue for quantum spin Hall case. This has important consequences since it has been widely communicated that non-local transport through edge channels in topological insulators  will have potential applications in low power information processing.
 \end{abstract}

\maketitle

1D edge modes are the hallmark of  quantum Hall(QH) and quantum spin Hall(QSH) set-ups\cite{sanvito,chulkov,Arjun,buti-sci}. These arise in quantum Hall case at high magnetic fields, however, in QSH case they arise at zero magnetic fields because of bulk spin orbit effects in 2D topological insulators\cite{sczhang}. QH edge modes are chiral while QSH edge modes are helical. In a remarkable  experiment conducted in Ref.[\onlinecite{koba}] and theoretically analyzed in Ref.[\onlinecite{seelig}],  an Aharonov-Bohm ring based four-probe set up was considered, in it was shown that the non-local resistance is less affected by dephasing than the local two-probe resistance. In this context we test whether in a QSH bar the non-local resistance will be adversely affected by  the twin effects of disorder and inelastic scattering the bane of any  phenomena which relies on complete quantum coherence.  In this work we show that non-local edge state transport in the QSH case is quite susceptible to disorder and even a single disordered probe can change the non-local resistance. Although it is well known that spin flip scattering adversely affects the non-local transport in QSH case\cite{jain}, we see in our work, in addition, that non spin flip scattering with disorder and inelastic scattering can greatly affect the non-local transport too. The reason for looking into this case is because of a point made in the abstract of Ref.[\onlinecite{Roth}]- that non-local transport through edge channels in topological insulators will have potential applications in low power information processing. We in this work show that this statement is not true in presence of disorder and/or inelastic scattering with or without spin flip processes.

\begin{figure*}
  \centering \subfigure[]{ \includegraphics[width=0.32\textwidth]{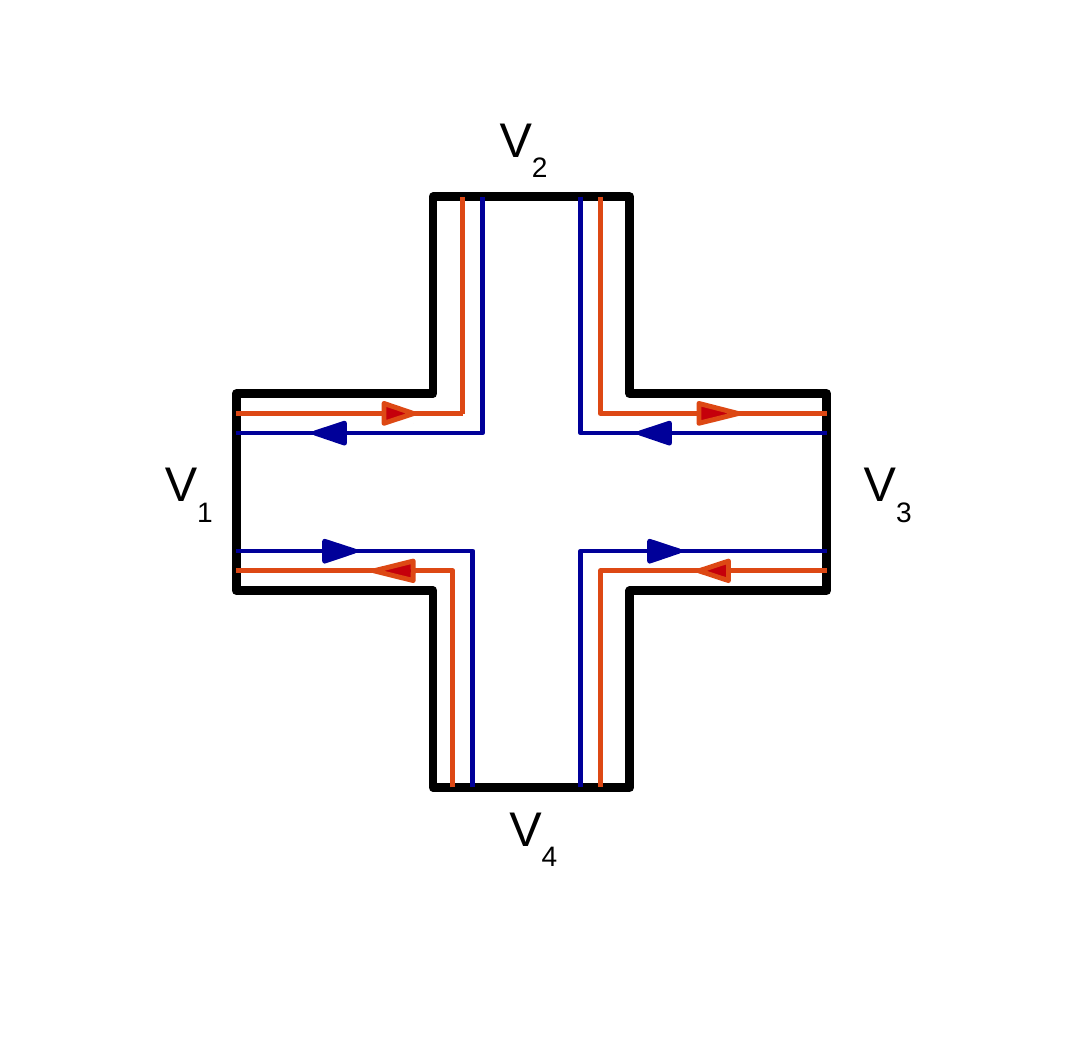}}
 \centering    \subfigure[]{ \includegraphics[width=.33\textwidth]{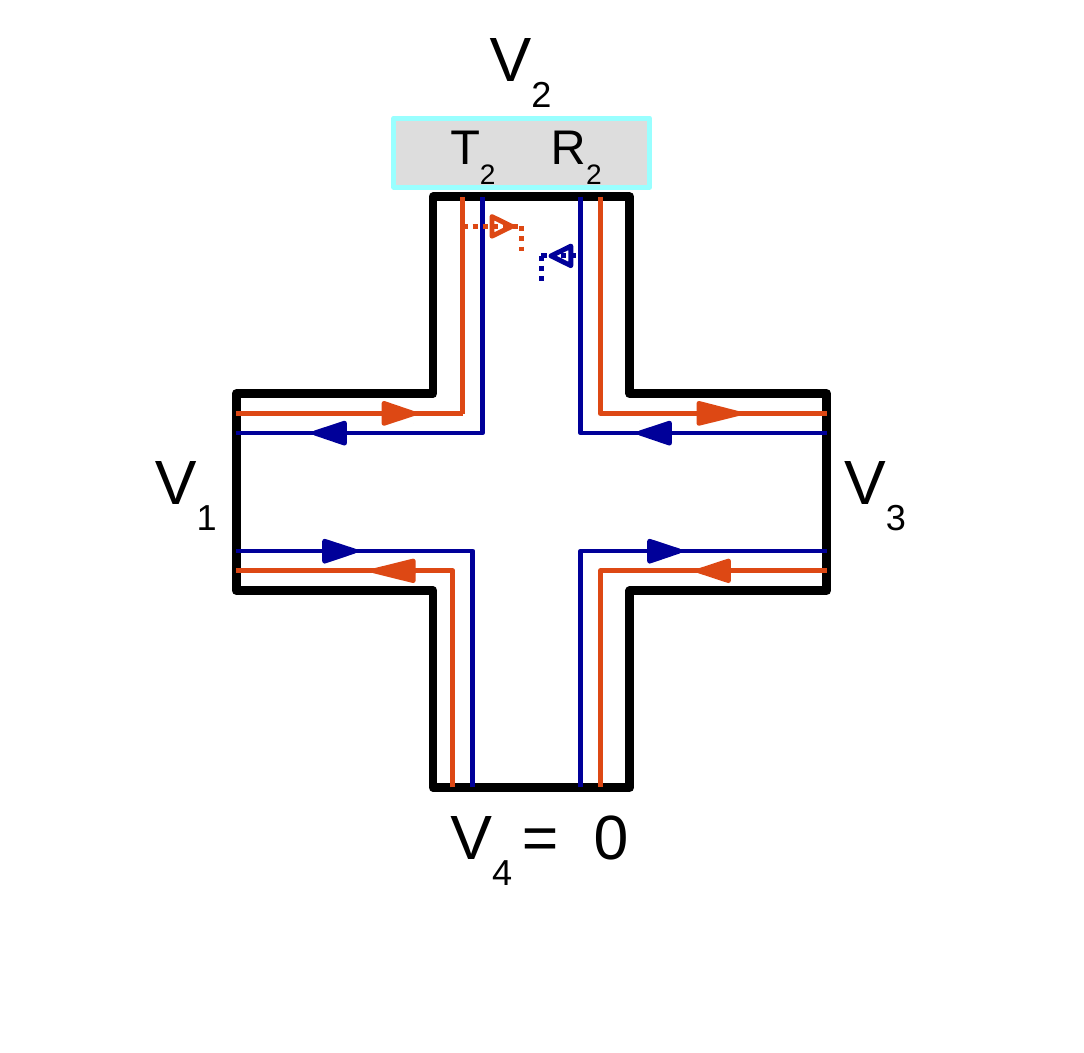}}
 \centering \subfigure[]{\includegraphics[width=.33 \textwidth]{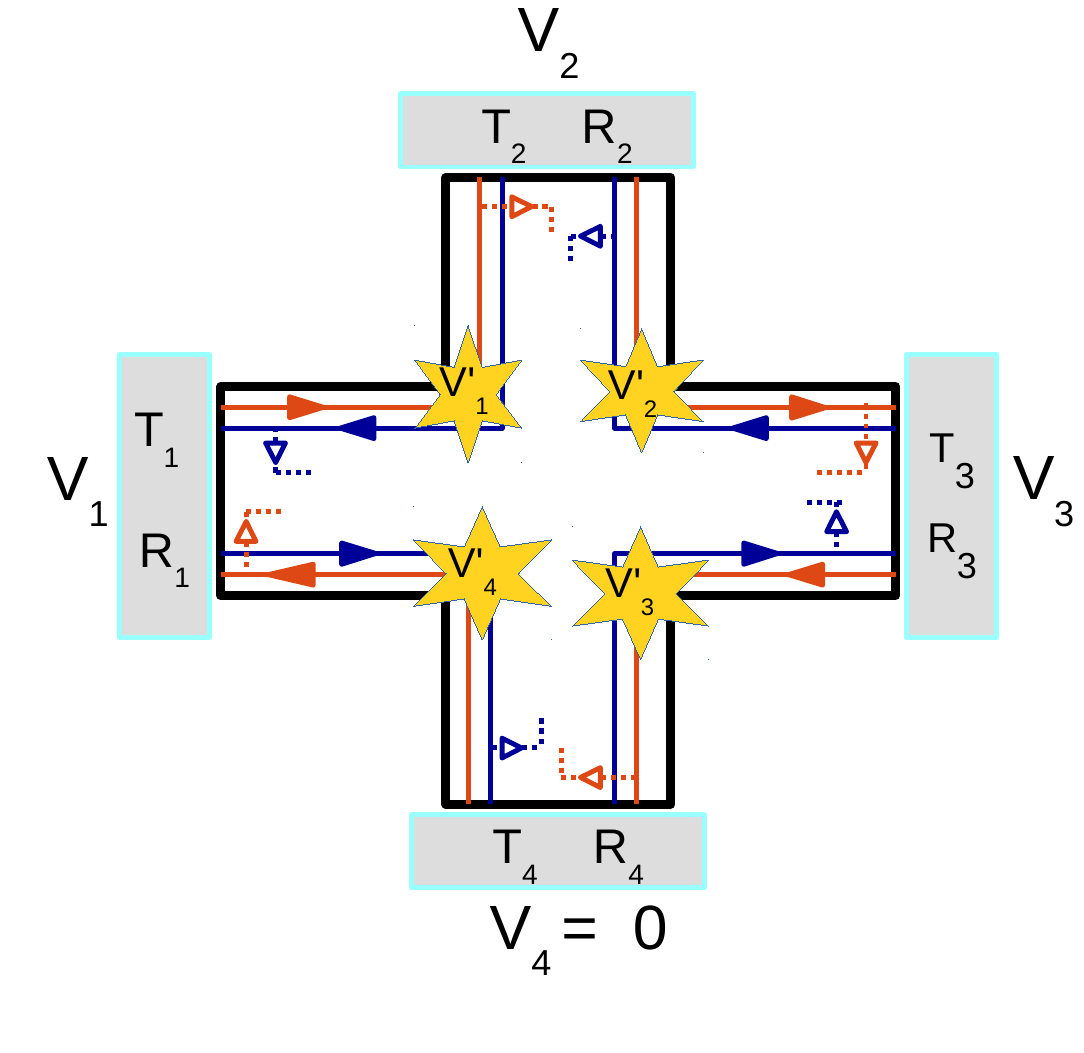}}
\vskip -0.2 in \caption{Four terminal Quantum Spin Hall bar showing QSH edge modes. These edge modes differ from their QH counterparts since these are spin polarized and helical. (a) Ideal case: absence of disorder and inelastic scattering, (b)Single disordered probe: $R_{1}=D_{1}M, T_{1}=(1-D_{1})M,$ represent the total reflection and transmission probability of edge modes from and into contact $1$ with the strength of disorder in contact 1 ranging from $0<D_{1}<1$. Dashed lines indicate the reflecting part of the edge modes due to disorder at contact 2, (c) All disordered contacts with inelastic scattering: Starry blobs indicate equilibration of contact potentials at those places }
\end{figure*} 

A disordered contact in contrast to an ideal contact-at which all electrons are transmitted, is one where some may be reflected back to the same contact. As sample size increases edge modes get affected by inelastic scattering too as long as inelastic scattering length $l_{in} < L$ (Length of sample). What inelastic scattering does is to equilibrate the populations and energies of electrons arising from two different contacts. This is what happens for inelastic scattering in QH samples but the situation changes for QSH samples. In QSH samples, equilibration can also happen between spin up and spin down edge modes via spin flip scattering. If spin flip scattering is absent, edge modes will still equilibrate due to inelastic scattering via electron-electron or electron-phonon interactions. However, in this case spin-up edge modes will equilibrate only with spin-up and not spin-down and similarly for spin-down edge modes. Edge states once equilibrated remain in equilibrium\cite{Arjun}. 

We analyze non-local vs local transport in a four probe QSH bar in five different cases-  a) ideal case-without any disorder or inelastic scattering, b) only single probe is disordered, c) all probes are disordered, d) all disordered probes in an inelastic set up with spin flip scattering and finally, e) all disordered probes in an inelastic set up without spin flip scattering. For completeness, we address the QH case too but as is well known there is no non-local edge state transport in QH systems since chiral non-local transport in QH sample is absent. Disorder and inelastic scattering do not change this. This phenomena is well known in the solid state and goes by the adage "You can not kill a dead horse"\cite{imry}. We therefore look for examples in non-local helical QSH transport and check whether disorder and inelastic scattering can result in deviations from what was seen in the experiment of Ref.[\onlinecite{koba}]. 

To study non-local transmission we use the Landauer-Buttiker(LB) formula\cite{buti,nikolajsen} which relates currents and voltages in a multi probe device. LB formula has been extended to 2D topological insulator with 1D QSH edge modes in Refs.\onlinecite{sanvito,chulkov,Arjun}:
\begin{equation}
I_{i}=\sum_{j} ( G_{ji} V_{i} - G_{ij} V_{j})=\frac{e^{2}}{h} \sum_{j=1}^{N} ( T_{ji} V_{i} - T_{ij} V_{j})
\end{equation}
 $V_{i}$ being the voltage at i$^{th}$ probe while $I_{i}$ is the current flowing from the same probe. $T_{ij}$ denotes the transmission from j$^{th}$ to i$^{th}$ probe with $G_{ij}$ being the associated conductance. N denotes the no. of probes/contacts in the system. In our case it is 4. In the QSH and QH samples we study in the various cases we consider the configuration wherein $1, 4$ are current probes and $2,3$ are voltage probes. 
\section{Ideal four probe QSH bar} 
The ideal case is depicted in Fig. 1(a). The conductance matrix relating the currents with voltages is given as follows:
\begin{equation}
G_{ij} =-\frac{e^{2} M}{h} \left( \begin{array}{cccc}
    -2 & 1& 0 & 1 \\
    1  &-2 & 1 & 0 \\ 0  & 1 & -2 & 1\\ 1  & 0 & 1 & -2 \\\end{array} \right),
\end{equation}

$M$ represents the total no. of modes for each spin. In set-ups as shown in Figure 1, M=1 for clarity. From Eqs. (1) and (2) and since $I_{2}=I_{3}=0$ as $2,3$ are voltage probes, further choosing $V_{4}=0$ (reference potential), we have $V_{3}=V_{2}/2=V_{1}/3$. Thus, non-local resistance $R_{NL}=R_{14,23}=\frac{V_2-V_3}{I_1}=\frac{h}{e^{2}M}\frac{1}{4}$ and local (two probe) resistance  $R_{2T}=R_{14,14}=\frac{h}{e^{2}M}\frac{3}{4}$.  This result is exactly what was obtained earlier in Ref. [\onlinecite{Roth}].

For QH ideal case in the similar way like QSH case we can get the conductance matrix relating the currents with voltages which is-
\begin{equation}
G_{ij} =-\frac{e^{2} M}{h} \left( \begin{array}{cccc}
    -1 & 0& 0 & 1 \\
    1  &-1 & 0 & 0 \\ 0  & 1 & -1 & 0\\ 0  & 0 & 1 & -1 \\\end{array} \right),
    \label{qh-ideal}
\end{equation}
Here, $M$ represents the total no. of modes and equating the current $I_2$, $I_3$ equal to zero, and the reference potential $V_4=0$, one can calculate $V_2=V_3=V_1$ which gives the non-local resistance- $R_{NL}=0$, and two probe resistance $R_{2T}=\frac{h}{e^2}\frac{1}{M}$.

\section{{Single disordered probe}} The case represented in Fig. 1(b), depicts a single disordered probe.
The conductance matrix relating the currents with voltages is given as follows:
\begin{equation}
G_{ij} =-\frac{e^{2} }{h} \left( \begin{array}{cccc}
    -2M & T_{2}& R_{2} & M \\
    T_{2} &-2T_{2} & T_{2} & 0 \\ R_{2}  & T_{2} & -2M & M \\ M & 0 & M & -2M \\\end{array} \right),
\end{equation}

From Eqs. (1) and (3), with $I_{2}= I_{3}=0$ and choosing as before $V_{4}=0$, we  get $R_{NL}=R_{14,23}=\frac{h}{2 e^{2} M}\frac{1}{2+D_{2}}$ and  $R_{2T}=R_{14,14}=\frac{h}{2 e^{2} M}\frac{3+D_{2}}{2+D_{2}}$, where $D_i, i=1-4$ are the strength of disorder at contact i and it is related to $R_i$ (total reflection probability at contact i) and $T_i$ (total transmission probability at contact i) by the relations- $R_i=D_i M$ and $T_i=(1-D_i)M$. In other words, $D_i$ is the reflection probability of each edge mode at contact `i'.  We see that both non-local and local (two probe) resistances deviate from their ideal values and are decreasing with disorder. This is in contrast to QH case in which they are independent of disorder.  In four probe QH bar with a single disordered probe at contact 2 with strength $D_2$,  the current coming out of probe 2 is $\frac{e^{2} }{h}T_2$ instead of $\frac{e^{2} M}{h}$ like in the ideal case and solving the similar conductance matrix for ideal QH case as in Eq. \ref{qh-ideal}, we will get $V_2=V_3=V_1$, which leads to the non-local resistance $R_{NL}=0$ as in ideal case, and the two terminal resistance $R_{2T}=\frac{h}{e^2}\frac{1}{M}$. 
\begin{figure*}
 \centering    \subfigure[Resistance vs. Disorder $D_4$ at current probe $4$, disorder in other contacts: $D_1 = 0.5$, $D_2 = D_3 = 0$ (Ideal voltage probes).]{ \includegraphics[width=.45\textwidth]{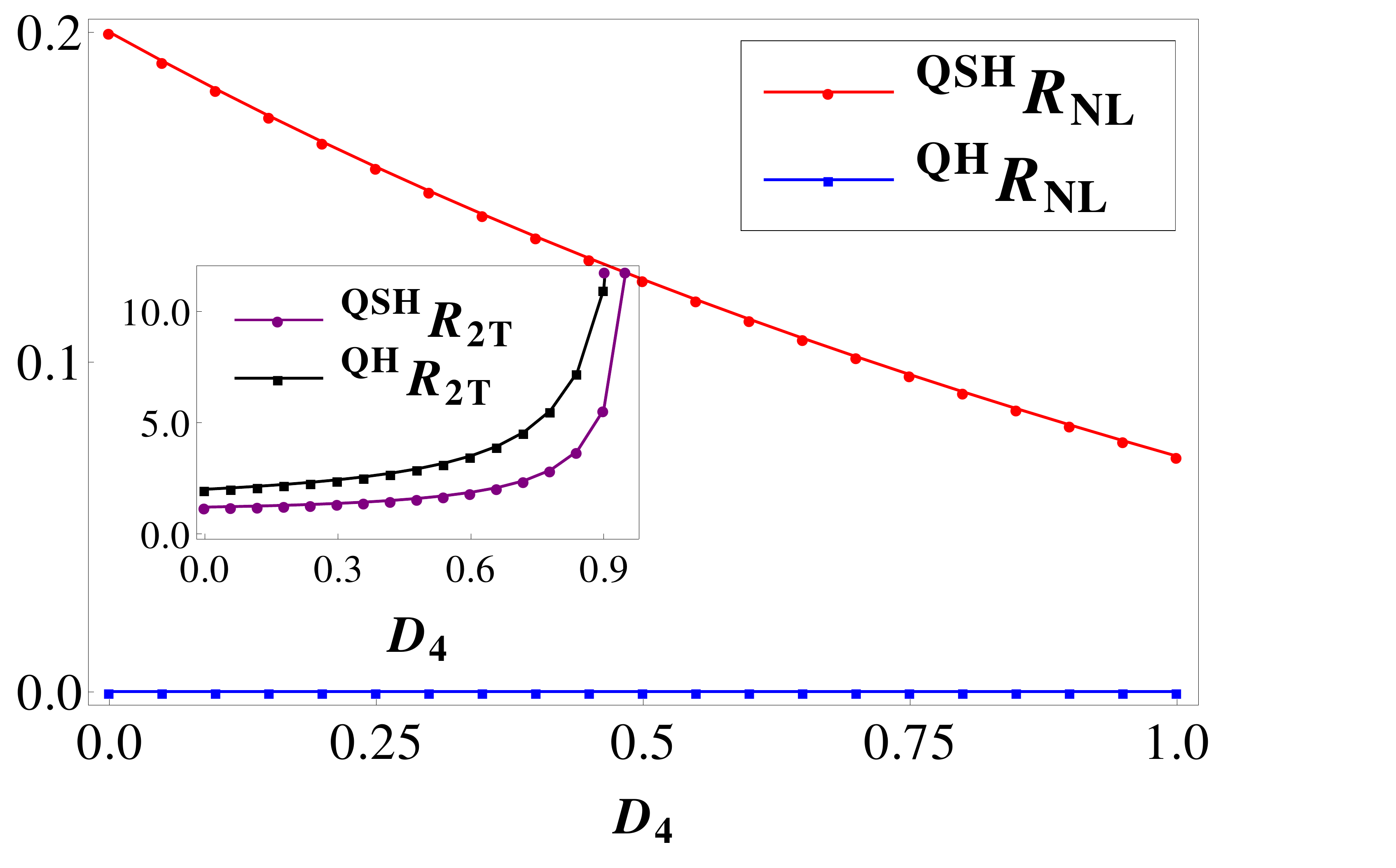}}
 \centering    \subfigure[Resistance  vs. Disorder $D_{2}$  at voltage probe $2$, disorder in other contacts are $D_{1}=D_{4}=0, D_{3}=0.5$ (ideal current probes). ]{ \includegraphics[width=.45\textwidth]{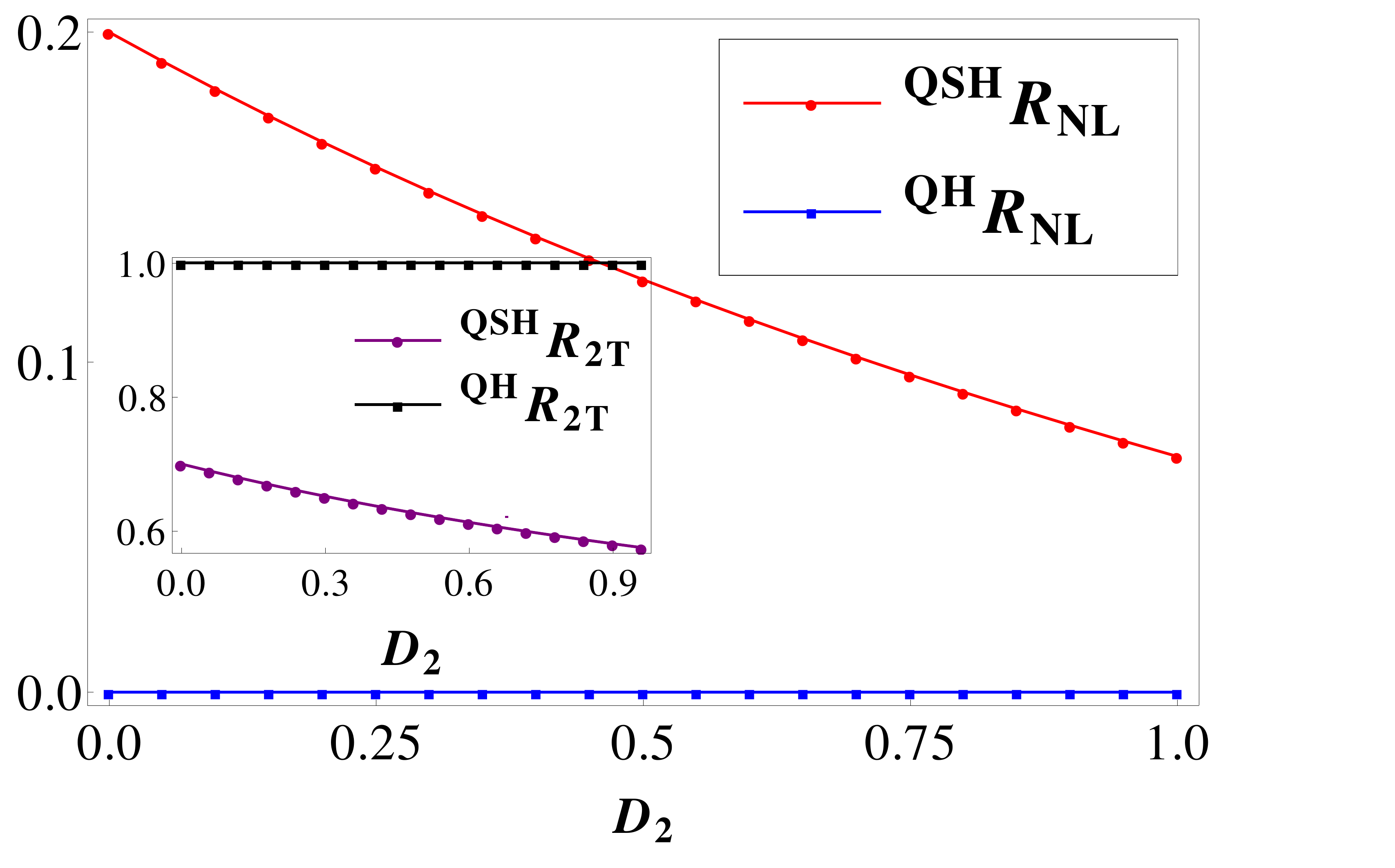}}
 \centering \subfigure[Resistance  vs. Disorder $D_{2}$ for the case when all probes are disordered, disorder in other contacts are $D_{1}=0.6, D_{3}=0.6, D_{4}=0.7$. ]{ \includegraphics[width=.45\textwidth]{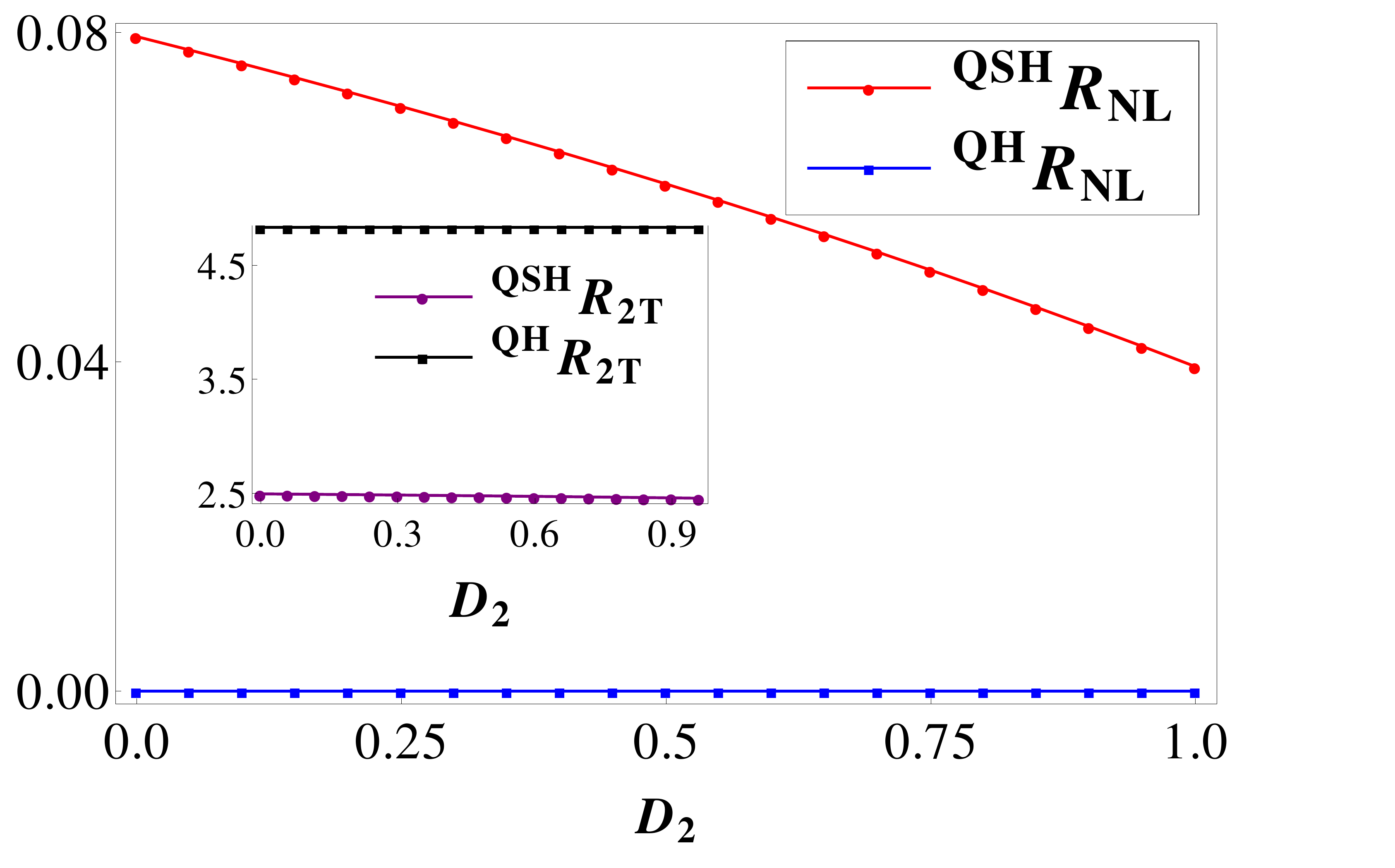}}
 \centering \subfigure[Resistance  vs. Disorder $D_{4}$ for the case when all probes are disordered, disorder in other contacts are $D_{1}=D_{2}=D_{3}=0.5$.]{ \includegraphics[width=.45\textwidth]{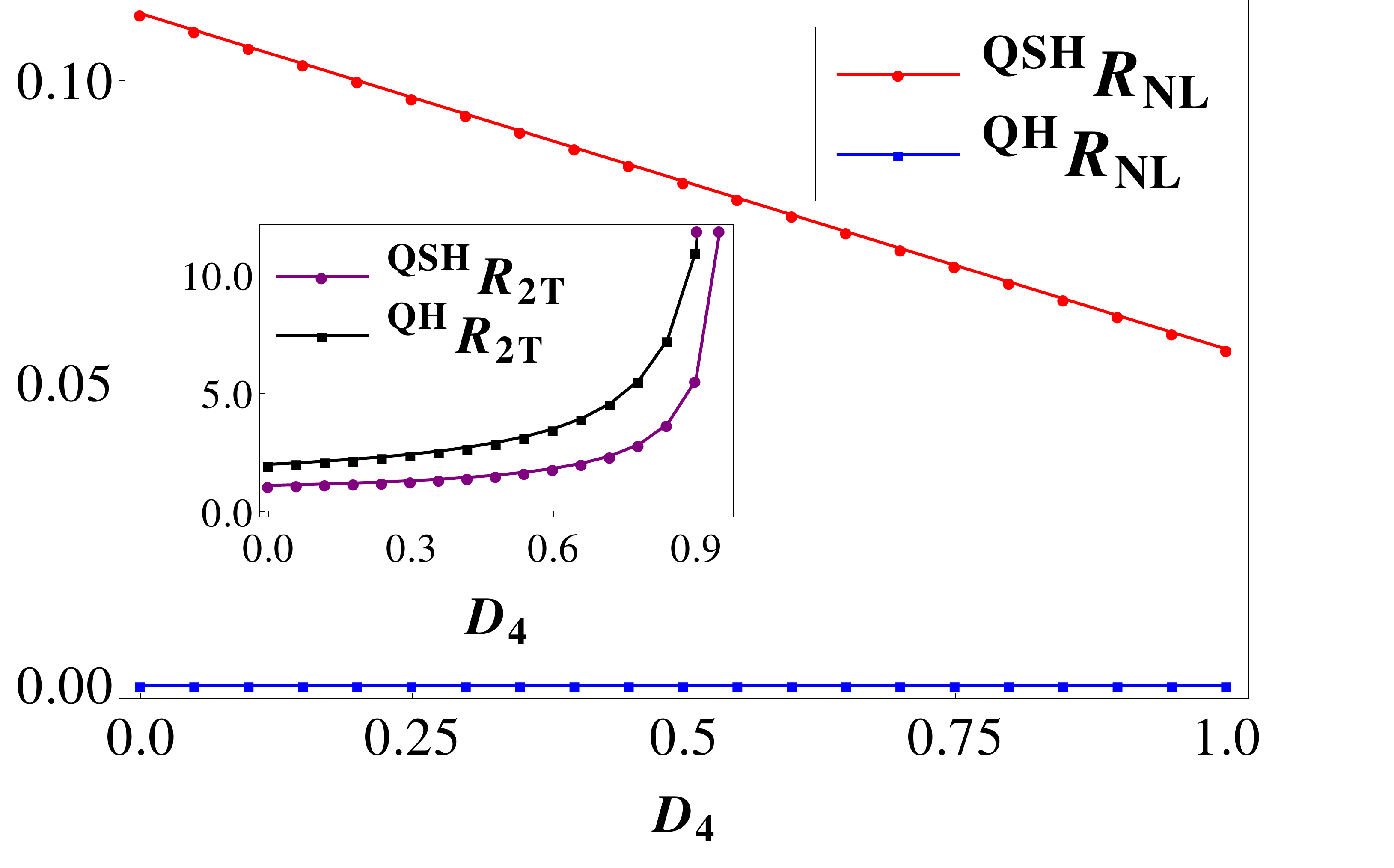}}
\caption{Non-local resistance $R_{NL} \mbox {    and local resistance }  R_{2T}$ in units of $\frac{h}{e^{2}}$  vs. Disorder. }
\end{figure*} 

\section{{Two or more disordered probes}}
 All contacts are considered to be disordered in general for this case. The conductance matrix relating the currents with voltages is given as follows:
\begin{equation}
G_{ij} =-\frac{e^{2} M}{h} \left( \begin{array}{cccc}
    -T^{11}  & T^{12} & T^{13} & T^{14} \\
    T^{21}  & -T^{22} & T^{23} & T^{24} \\ 
    T^{31}  & T^{32} & -T^{33} & T^{34} \\
     T^{41}  & T^{42} & T^{43} & -T^{44} \\ \end{array} \right)
\end{equation}
M represents the total no. of modes. In set up as shown in Fig. 1c, only one mode is shown. 
In conductance matrix (4), $T^{13}$-the total transmission probability from contact $3$ to 
$1$, can be written as- \[T^{13}=\frac{[(1-D_{3})D_{4}(1-D_{1})+(1-D_{3})D_{2}(1-D_{1})]M}{1-D_{1}D_{2}D_{3}D_{4}},\] The first term in the numerator of $T^{13}$ is  due to the spin up edge mode and second term is due to the spin down edge mode. Lets explain the terms in the numerator.  In the first term electron coming out of contact 3 with probability $(1-D_3)$,  is reflected at contact 4 with probability $D_4$ and then it transmits at contact 1 with probability $(1-D_1)$, so the probability of transmission for spin up electrons  is $(1-D_{3})D_{4}(1-D_{1})M$. But this is the shortest possible path among the many other path from contact 3 to 1. So to get the total transmission probability for spin up electrons we need to sum over all the paths from contact 3 to 1. For the second path for spin up electron, suppose an electron coming out of contact 3 with probability $(1-D_3)$, then it reflects at contact 4 with probability $D_4$ again   it is reflected from contact 1 with $D_1$, from contact 2 with $D_2$,  from contact 3 with $D_3$, from contact 4 with $D_4$ and then finally it transmits to contact 1 with probability $1-D_1$. So the second path becomes $(1-D_3)D_4^2D_1D_2D_3(1-D_1)M$. So when we sum over all the paths we get a factor $(1-D_1D_2D_3D_4)$ in the denominator. So  finally transmission probability for spin up electron is 
$\frac{[(1-D_{3})D_{4}(1-D_{1})]M}{1-D_{1}D_{2}D_{3}D_{4}}$. 
Similarly for spin down electron from contact 3 to 1 the transmission probability is $\frac{[((1-D_{3})D_{2}(1-D_{1})]M}{1-D_{1}D_{2}D_{3}D_{4}}$. In this way we get $T^{13}$, and can calculate the other matrix elements also. The non-local and local resistances are  thus in units of $\frac{h}{e^{2}}$-

\begin{eqnarray}
R_{NL}&=&\frac{(1-D_2D_3)(1-D_1D_4)}{Den},\\
R_{2T}&=&\frac{(1-D_1D_4)Num}{(1-D_1)(1-D_4)Den}
\end{eqnarray}
{\small
\begin{eqnarray*}
Num&=&3+D_1D_4+D_3(1-D_1D_4)+D_2(1-D_3-D_1D_4-3D_1D_3D_4)\\
Den&=&2 + D_2 + D_3 + D_4 - D_2 D_3 D_4 - D_1 (-1 + D_3 D_4 + D_2 (D_3 + D_4 + 2 D_3 D_4))
\end{eqnarray*}
}
 From Figs. 2(c),(d) it is evident that local resistance for QSH case can increase or decrease with increasing disorder (depending on the current probe disorder or voltage probe disorder respectively), but the non-local resistance always decrease with increasing disorder. This is in contrast to the non-local QH case which again does not deviate from the ideal result. In QH case, for all probe disorder and all probe disorder with inelastic scattering we get identical results- $R_{NL}=0$, and $R_{2T}=\frac{h}{e^2}\frac{(1-D_1D_4)}{(1-D_1)(1-D_4)M}$, where $D_1, D_4$ denote the disorder strengths at probe 1 and 4 respectively. Only difference taking into play between QH and QSH edge modes is that the first one is chiral and the second one is helical, and due to this helical (two opposite chiralities) edge modes the non-local resistance in QSH is affected by disorder while it remains unaffected in QH case. For all disordered probes  with inelastic scattering included- the condition $V_2=V_3$ still holds for the QH system which is due to the chirality of edge modes, leading to vanishing non-local resistance, this is easy to understand-  two adjacent voltage probes   (in between there is no current probe) will always have same potential for chiral edge modes, leading to zero non-local resistance irrespective of any  disorder at the probes.

\section{{All disordered probes with inelastic and spin-flip scattering}} We now address the case of QSH edge modes in presence of completely disordered contacts with inelastic scattering included. Further, we also take into consideration the fact that an electron can flip its spin. In Figure 1(c), the length between disordered contacts is taken to be larger than inelastic scattering length. If an inelastic scattering event occurs, edge modes originating from different probes with different energies are equilibrated to a common potential\cite{Arjun}. 

An important question might arise regarding the fact that inelastic scattering in Fig. 1(c) seems to occur only at a specific point, this is a misleading conclusion from Fig. 1(c),  for reasons to do with clarity we have put the starry blob in Fig. 1(c)  at a specific place. However,   equilibration between electrons in edge modes emanating from contacts 1 and 2  can happen anywhere between contacts 1 and 2. Once an electron is inelastically scattered it remains inelastically scattered. Inelastic scattering does not happen only at a single point, it can happen anywhere.  It is a single energy to which electrons are inelastically scattered the equilibration potential (primed values for the $V_{i}$'s in Fig. 1(c)). 

In Figure 1(c), one sees that electrons coming from probes 1 and 2 are equilibrated to potential $V_{1}^{\prime}$- indicated in Figure 1(c) as a starry blob to indicate equilibration of electrons coming from different probes at different potentials (unprimed values) to a new potential (the primed values) via inelastic scattering. If probes 1 and 4 are chosen to be current probes then no current flows into the other voltage probes $2\&3$. Say a current $\frac{e^2}{h} (T_{2} V_{1}^{\prime}+T_{2} V_{2}^{\prime})$ enters probe 2, the part $\frac{e^2}{h} T_{2} V_{1}^{\prime}$ is the spin-up contribution while the  part  $\frac{e^2}{h} T_{2} V_{2}^{\prime}$ is the spin-down one, which moves in exactly the opposite direction. Further, since  current $\frac{e^2}{h} 2 T_{2} V_{2}$ leaves voltage probe 2 and as net current through probe 2 is zero, this results in $V_{2}=(V_{1}^{\prime}+V_{2}^{\prime})/2$. Similarly for voltage probe 3,  $V_{3}=(V_{2}^{\prime}+V_{3}^{\prime})/2$.

Writing the current voltage relations for each probe separately, and eschewing the previous way of writing it in matrix form as there are not only 4 potentials $V_{1}$ to $V_{4}$, we also have the equilibrated potentials $V_{1}^{\prime}$ to $V_{4}^{\prime}$, we get-.
\begin{eqnarray}
I_{1}&=& T_{1}(2V_{1}-V_{1}^{\prime}-V_{4}^{\prime}),\nonumber\\
I_{2}&=& T_{2}(2V_{2}-V_{1}^{\prime}-V_{2}^{\prime}),\nonumber\\
I_{3}&=& T_{3}(2V_{3}-V_{2}^{\prime}-V_{3}^{\prime}),\nonumber\\
I_{4}&=& T_{4}(2V_{4}-V_{3}^{\prime}-V_{4}^{\prime}).
\end{eqnarray}
The condition that net current into voltage probes $2,3$ is zero begets the following relations between the probe potentials: $V_{2}=(V_{1}^{\prime}+V_{2}^{\prime})/2 $ and $V_{3}=(V_{2}^{\prime}+V_{3}^{\prime})/2$. Net spin-up current out of probe 2 is $\frac{e^2}{h} (T_{2} V_{2}^{}+R_{2} V_{1}^{\prime})$ while the net spin-down current out of probe 3 is $\frac{e^2}{h} (T_{3} V_{3}^{}+R_{3} V_{3}^{\prime})$ and their sum is equal to $\frac{e^2}{h} 2 M V_{2}^{\prime}$ which is the net current at voltage $V_{2}^{\prime}$- the equilibrated potential due to inelastic scattering between probes 2 and 3. Since there are 4 equilibrated potentials we will have four such conditions. The origin of the first condition has already been explained, we  list the rest here:
\begin{eqnarray}
\frac{e^2}{h} (T_{2} V_{2}^{}+R_{2} V_{1}^{\prime}) + \frac{e^2}{h} (T_{3} V_{3}^{}+R_{3} V_{3}^{\prime}) &=&\frac{e^2}{h} 2 M  V_{2}^{\prime}\nonumber\\
\frac{e^2}{h} (T_{1} V_{1}^{}+R_{1} V_{4}^{\prime}) + \frac{e^2}{h} (T_{2} V_{2}^{}+R_{2} V_{2}^{\prime}) &=&\frac{e^2}{h} 2 M  V_{1}^{\prime}\nonumber\\
\frac{e^2}{h} (T_{3} V_{3}^{}+R_{3} V_{2}^{\prime} )+ \frac{e^2}{h} (T_{4} V_{4}^{}+R_{4} V_{4}^{\prime})&=&\frac{e^2}{h} 2 M  V_{3}^{\prime}\nonumber\\
\frac{e^2}{h} (T_{4} V_{4}^{}+R_{4} V_{3}^{\prime} )+ \frac{e^2}{h} (T_{1} V_{1}^{}+R_{1} V_{1}^{\prime})&=&\frac{e^2}{h} 2 M  V_{4}^{\prime}.
\end{eqnarray}

Solving Eq. (8), leads to a relation between the equilibrated potentials $V_{i}^{\prime}, i=1,..4$ in terms of the probe potentials  $V_{i}^{}, i=1,..4$. Replacing the obtained $V_{i}^{\prime}, i=1,..4$ in Eq. ~5, gives the necessary resistances. We plot in Fig. 3, the non-local and local resistances. The erratic behaviour for the non-local QSH resistance comes out clearly. For one set of disorder parameters the non-local resistance decreases while for another set it monotonically increases as function of disorder. The local resistance on the other hand monotonically increases with increasing disorder.
 
\begin{figure}
 \centering    \subfigure[Resistance  vs. Disorder $D_{2}$ in voltage probe for inelastic scattering with spin-flip, disorder in other contacts are $D_{1}=0.5, D_{3}=0.33, D_{4}=0.25$]{ \includegraphics[width=.45\textwidth]{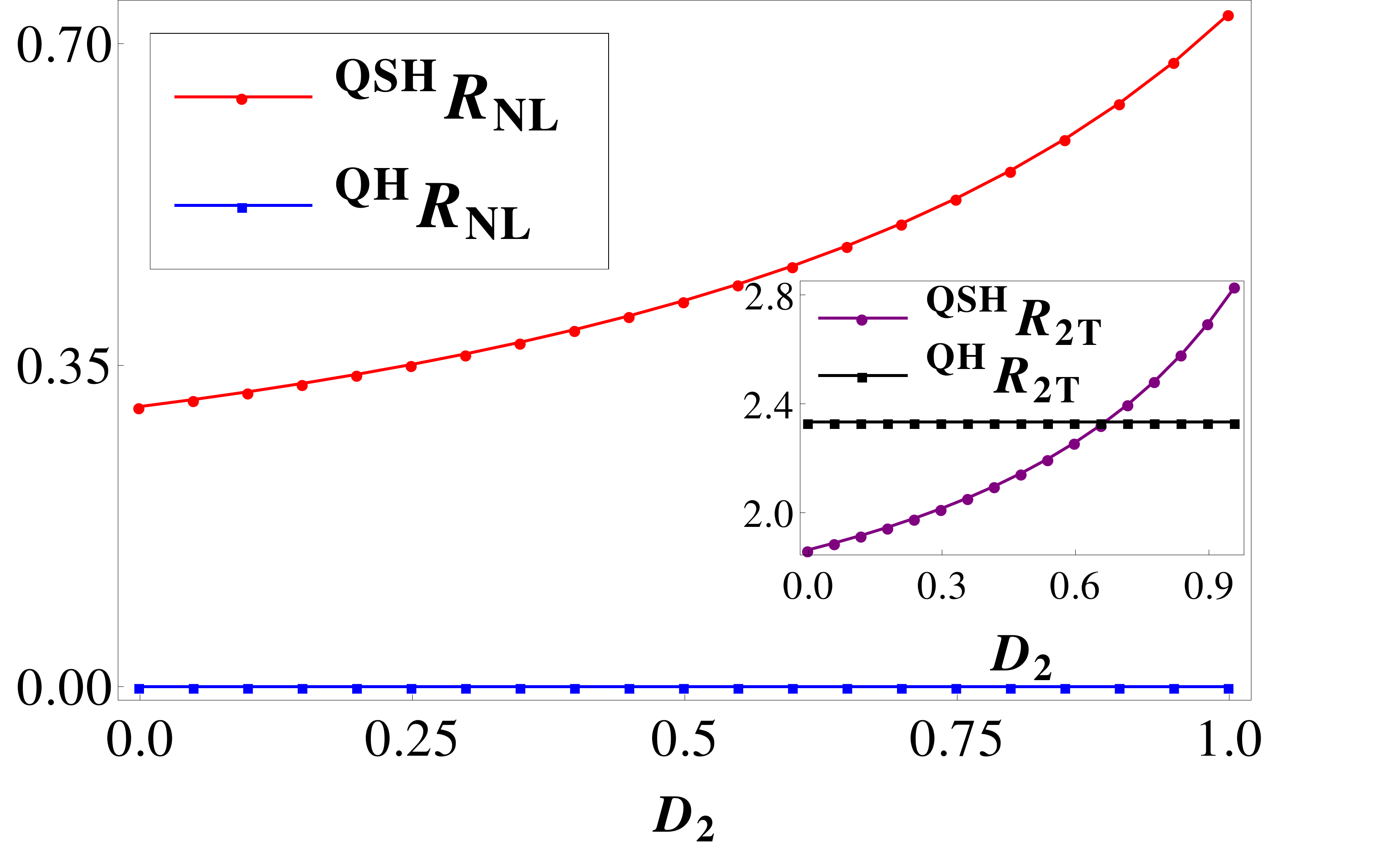}}
 \centering    \subfigure[Resistance  vs. Disorder $D_{4}$ in current probe  for inelastic scattering with spin-flip, disorder in other contacts are $D_{1}=0.5, D_{2}=0.25, D_{3}=0.33$]{ \includegraphics[width=.45\textwidth]{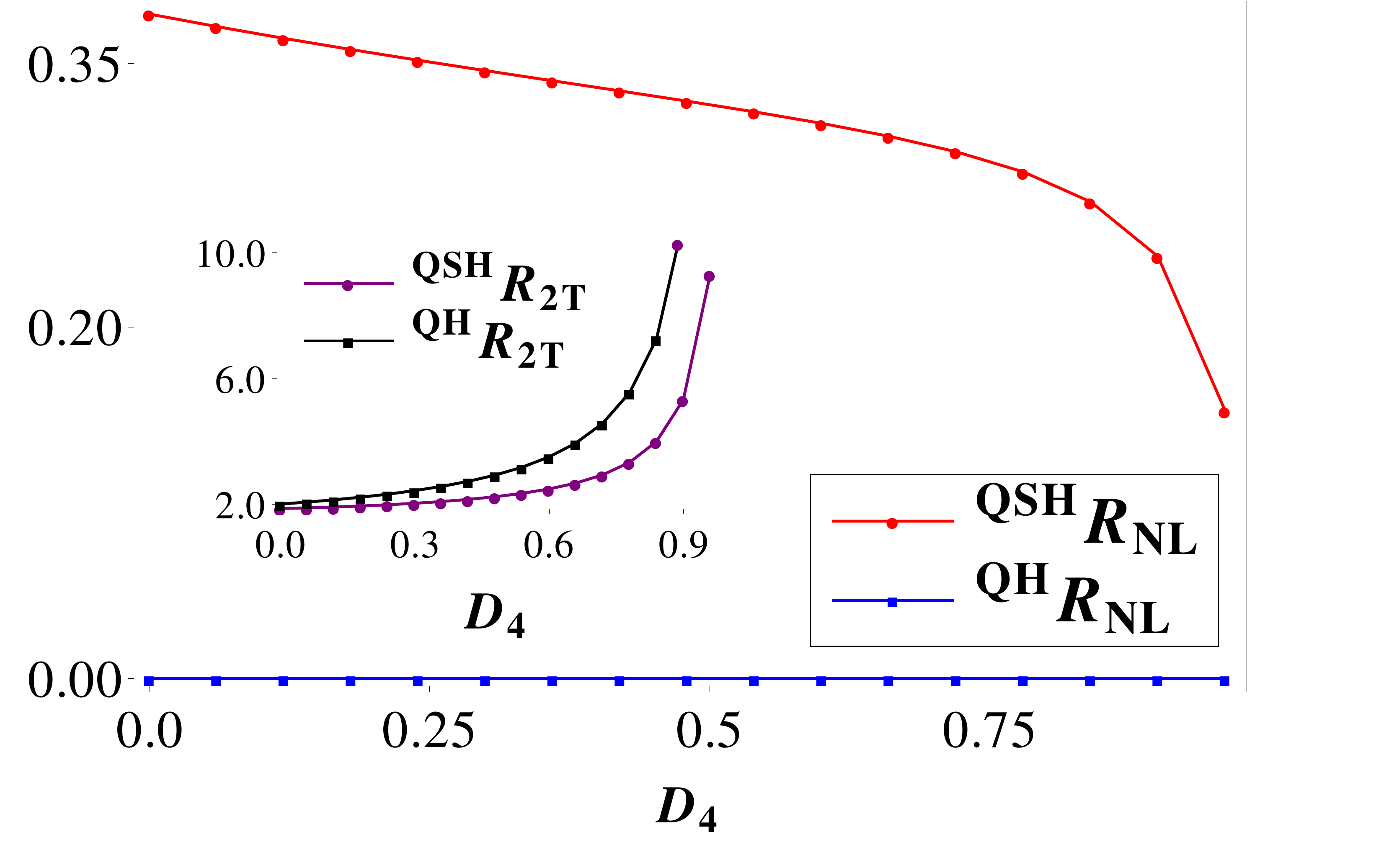}}
\caption{Non-local resistance $R_{NL}, \mbox{  and local resistance } R_{2T}$ in units of $\frac{h}{e^{2}}$ vs. Disorder. }
\end{figure} 

\section{{All disordered probes with inelastic but without spin-flip scattering}}
 Finally, we deal with the case of QSH edge modes with all disordered probes including inelastic scattering, however in absence of any spin-flip scattering. We now only have equilibration between same spin edge modes. In Fig. 4(a), spin-up electrons originating in probes 1 and 4 are equilibrated to potential $V_{1}^{\prime}$ while spin-down electrons coming from probes 2 and 3 are equilibrated to potential $V_{1}^{\prime \prime}$. The potentials $V_{i}^{\prime}, i=2 \mbox{ - } 4$ denote equilibration of spin-up edge modes while the potentials $V_{i}^{\prime\prime}, i=2\mbox{-} 4$ denote equilibration of spin-down edge modes. Since probes 1 and 4 are current probes and no current flows into the voltage probes, say a spin-up current $\frac{e^2}{h} (T_{2} V_{1}^{\prime}$), and a spin down current $\frac{e^2}{h} (T_{2} V_{2}^{\prime\prime})$ enter probe 2, while current $\frac{e^2}{h} 2 T_{2} V_{2}$ leaves probe 2 (voltage probe with net current through it zero), giving $V_{2}=(V_{1}^{\prime}+V_{2}^{\prime\prime})/2$, a similar thing happens at probe 3.

 There are now not only the 4 potentials $V_{1} \mbox {-} V_{4}$, we also have the equilibrated spin-up potentials $V_{1}^{\prime} \mbox{-} V_{4}^{\prime}$ and the spin down potentials $V_{1}^{\prime\prime} \mbox{-} V_{4}^{\prime\prime}$. We write the current voltage relations in this condition as follows:
\begin{figure}[t]
\centering \subfigure[All disordered contacts with inelastic scattering but without any spin-flip scattering]{\includegraphics[width=.4\textwidth] {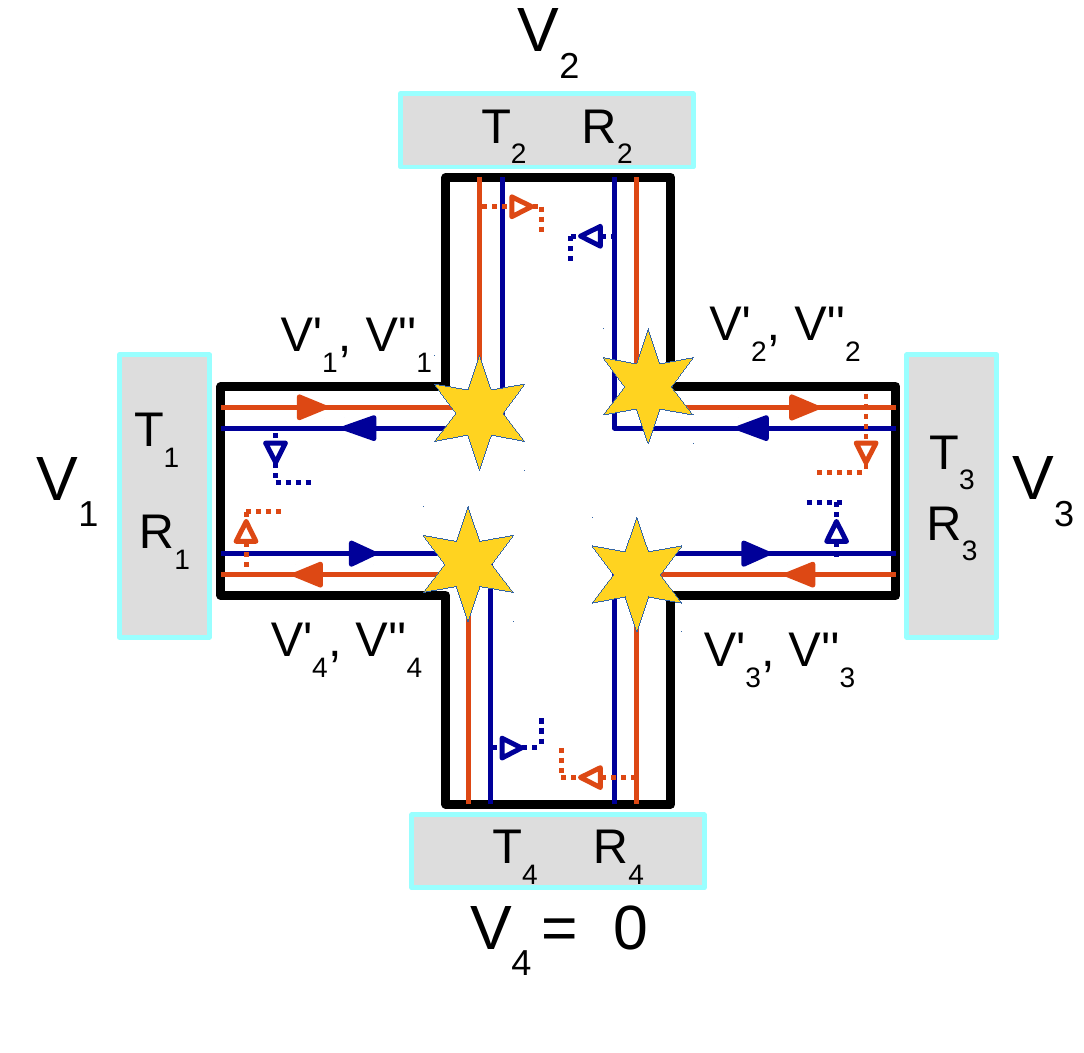}}
 \centering    \subfigure[Resistance  vs. Disorder $D_{2}$ for inelastic scattering without spin-flip, parameters are $D_{1}=0.5, D_{3}=0.33, D_{4}=0.25$] {\includegraphics[width=.4\textwidth]{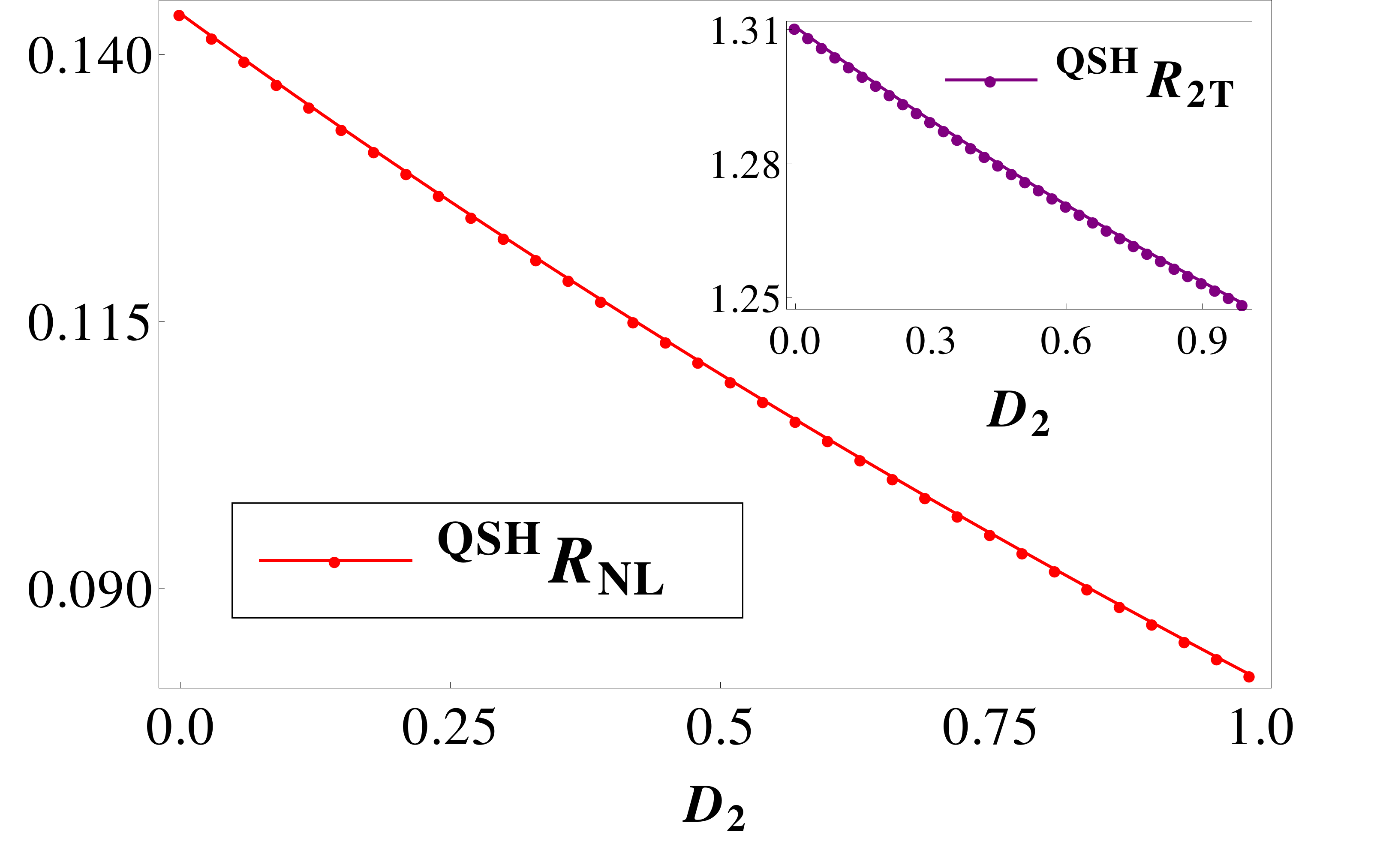}} \centering    \subfigure[Resistance  vs. Disorder $D_{2}$ for inelastic scattering with spin-flip, same parameters as in (b).] {\includegraphics[width=.4\textwidth]{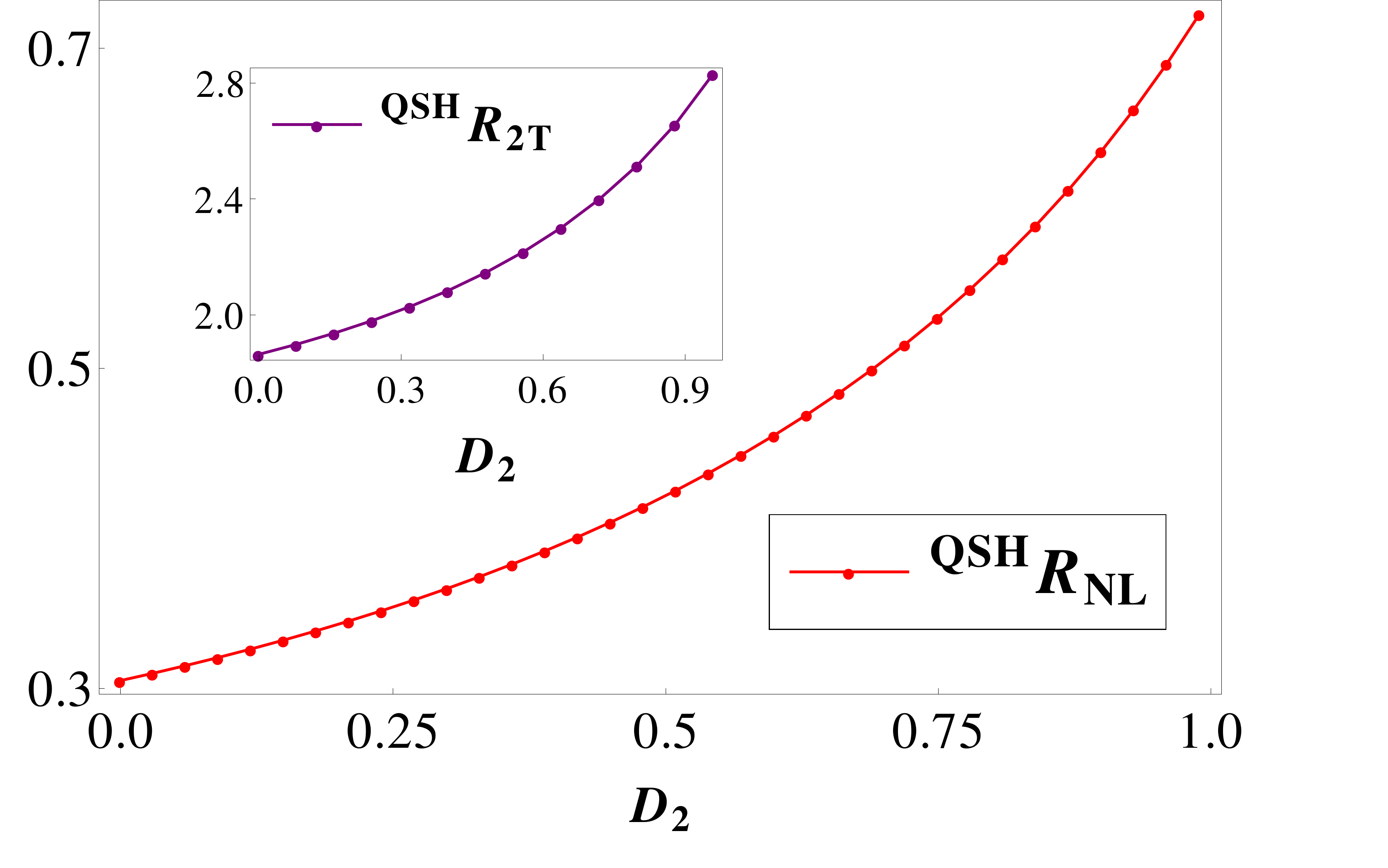}}
\caption{Non-local resistance $R_{NL}, \mbox{   and local resistance }  R_{2T}$ in units of $\frac{h}{e^{2}}$ vs. Disorder for with inelastic scattering (b) without spin-flip and (c) with spin-flip.}
\end{figure} 
\begin{eqnarray}
I_{1}&=&\frac{e^2}{h}  T_{1}(2V_{1}-V_{1}^{\prime\prime}-V_{4}^{\prime}),\nonumber\\
I_{2}&=&\frac{e^2}{h}  T_{2}(2V_{2}-V_{2}^{\prime\prime}-V_{1}^{\prime}),\nonumber\\
I_{3}&=&\frac{e^2}{h}  T_{3}(2V_{3}-V_{3}^{\prime\prime}-V_{2}^{\prime}),\nonumber\\
I_{4}&=&\frac{e^2}{h}  T_{4}(2V_{4}-V_{4}^{\prime\prime}-V_{3}^{\prime}).
\end{eqnarray}
 When the condition of net current into voltage probes $2,3$ is zero is enforced we get the following relations between probe potentials: $V_{2}=(V_{1}^{\prime}+V_{2}^{\prime\prime})/2, V_{3}=(V_{2}^{\prime}+V_{3}^{\prime\prime})/2$. Equilibration of net spin-up current from probe 1, which is the sum  $\frac{e^2}{h} (T_{1} V_{1}^{}+R_{1} V_{4}^{\prime})$, will be equal to $\frac{e^2}{h} M V_{1}^{\prime}$ the net current out of the spin-up equilibrated potential $V_{1}^{\prime}$. Similarly, the net spin up currents out of probes 2-4 are equilibrated to the potentials $V_{i}^{\prime}, i=2-4$ as in Eq.~8. 
We follow the same procedure for the down spin currents and these are written below in Eq.~8. The origin of the first equation has already been explained above herein below we  list all of them:
\begin{eqnarray}
{\frac{e^2}{h} (T_{1} V_{1}^{}+R_{1} V_{4}^{\prime})=\frac{e^2}{h} M  V_{1}^{\prime}}\nonumber     &,&   {\frac{e^2}{h} (T_{1} V_{1}^{}+R_{1} V_{1}^{\prime\prime})=\frac{e^2}{h}MV_{4}^{\prime\prime}},\nonumber \\
{\frac{e^2}{h} (T_{2} V_{2}^{}+R_{2} V_{1}^{\prime})=\frac{e^2}{h}M  V_{2}^{\prime}}  \nonumber &,& {\frac{e^2}{h} (T_{2} V_{2}^{}+R_{2} V_{2}^{\prime\prime})=\frac{e^2}{h}MV_{1}^{\prime\prime}},\nonumber\\
{\frac{e^2}{h} (T_{3} V_{3}^{}+R_{3} V_{2}^{\prime})=\frac{e^2}{h}M  V_{3}^{\prime}} \nonumber &,&  {\frac{e^2}{h} (T_{3} V_{3}^{}+R_{3} V_{3}^{\prime\prime})=\frac{e^2}{h}MV_{2}^{\prime\prime}},\nonumber\\
{\frac{e^2}{h} (T_{4} V_{4}^{}+R_{4} V_{3}^{\prime})=\frac{e^2}{h}M  V_{4}^{\prime}}  &,&    {\frac{e^2}{h} (T_{4} V_{4}^{}+R_{4} V_{4}^{\prime\prime})=\frac{e^2}{h}MV_{3}^{\prime\prime}}.
\end{eqnarray}
Solving these eight equations, gives the equilibrated potentials $V_{i}^{\prime} \mbox { and } V_{i}^{\prime\prime}, i=1-4$ in terms of the probe potentials  $V_{i}^{}, i=1-4$. The obtained $V_{i}^{\prime} \mbox{and } V_{i}^{\prime\prime}, i=1-4$ are replaced in Eq. ~8, and we derive the necessary resistances.  Due to the length of the expressions for resistances we refrain from explicitly writing them but analyze them via Figure 4.
%We have-
%\begin{eqnarray}
%R_{NL}=R_{14,23}&=&\frac{h}{2 e^{2} M}\frac{(1-D_{2}D_{3})(1-D_{1}D_{4})}{Den}, \nonumber\\ 
%R_{2T}=R_{14,14}&=&\frac{h}{2 e^{2}M}\frac{(-1+D_{1}D_{4})(Num)}{(1-D_{1})(1-D_{4})(Den)}.\nonumber\\
%Den&=&2+D_{2}+D_{3}+D_{4}-D_{2}D_{3}D_{4}-\nonumber\\
%&-&D_{1}(-1+D_{3}D_{4}+D_{2}(D_{3}+D_{4}+2D_{3}D_{4}))\nonumber\\
%Num&=&(-3-D_{1}D_{4}+D_{3}(-1+D_{1}D_{4})\nonumber\\
%&+&D_{2}(-1+D_{3}+D_{1}D_{4}+3D_{1}D_{3}D_{4})
%\end{eqnarray}
 
In Figure 4 (c) we plot the non-local and 2-Terminal resistances for QSH case with spin-flip and in 4(b) we plot the same without spin-flip. We see that the non-local resistance completely changes when one discards spin flip scattering as in Figure 4(b). The non-local resistance and 2-terminal resistance decrease monotonically for without spin flip while increase monotonically with spin flip. To conclude this section, spin-flip scattering has a non-trivial effect on the non-local response while disorder and inelastic scattering lead to a monotonic reduction in the non-local resistance for QSH edge modes, spin-flip can enhance the non-local resistance too.

\section{{Conclusion}} To conclude we establish here that non-local transport in case of  helical QSH edge modes are not less impervious to dephasing as previously understood in general non-helical four-terminal Aharonov-Bohm type set ups\cite{koba,seelig}. Further, they are modified quite drastically by disorder, even a single disordered probe reduces significantly the non-local resistance. This puts a big question mark over the usefulness of non-local QSH transport in low power information processing as reported in several works, see Ref. [\onlinecite{Roth}]. At the end we summarize the results of our work in the following table.
\begin{widetext}

{
\begin{center}
\begin{tabular}{ |p{3.4cm}|p{1.5cm}|p{1.5cm}|p{1.5cm}|p{1.5cm}|p{1.5cm}|p{1.5cm}|p{1.5cm}|p{1.5cm}|} 
 \hline
& \multicolumn{4}{|c|}{QH Edge Modes}  & \multicolumn{4}{|c|}{QSH Edge Modes}  \\ 
\hline
& \multicolumn{2}{|c|}{$\mbox{  \hskip 0.5 cm}R_{NL}\mbox{  \hskip 0.5 cm}$}  & \multicolumn{2}{|c|}{$R_{2T}$}&\multicolumn{2}{|c|}{$R_{NL}$}& \multicolumn{2}{|c|}{$R_{2T}$} \\ 
\hline
 %&Current Probe Disorder & Voltage Probe Disorder&Current Probe Disorder & Voltage Probe Disorder&Current Probe Disorder & Voltage Probe Disorder&Current Probe Disorder & Voltage Probe Disorder\\ 
\hline
Ideal Case&\multicolumn{2}{|c|}{0}&\multicolumn{2}{|c|}{$\frac{h}{e^2}\frac{1}{M}$}&\multicolumn{2}{|c|}{$\frac{h}{e^2M}\frac{1}{4}$}&\multicolumn{2}{|c|}{$\frac{h}{e^2M}\frac{3}{4}$}\\[2ex]
\hline
Single disordered probe&\multicolumn{2}{|c|}{0}&\multicolumn{2}{|c|}{$\frac{h}{e^2}\frac{1}{M}$}&\multicolumn{2}{|c|}{$\frac{h}{2 e^{2} M}\frac{1}{2+D_{2}}$}&\multicolumn{2}{|c|}{$\frac{h}{2 e^{2} M}\frac{3+D_{2}}{2+D_{2}}$}\\
\hline
Two or more  disordered probes&\multicolumn{2}{|c|}{0}&\multicolumn{2}{|c|}{$\frac{h}{e^2}\frac{(1-D_1D_4)}{(1-D_1)(1-D_4)M}$}&\multicolumn{2}{|c|}{Disorder dependent (Fig.~ 2)}&\multicolumn{2}{|c|}{Disorder dependent (Fig.~2)}\\
\hline
Disorder+inelastic scattering (with spin flip)&\multicolumn{4}{|c|}{Not applicable}&\multicolumn{2}{|c|}{Disorder dependent (Fig.~3)}&\multicolumn{2}{|c|}{Disorder dependent (Fig.~ 3)}\\
\hline
Disorder+inelastic scattering (without spin flip)&\multicolumn{2}{|c|}{0}&\multicolumn{2}{|c|}{Disorder dependent (Fig.~3)}&\multicolumn{2}{|c|}{Disorder dependent (Fig.~4 (b))}&\multicolumn{2}{|c|}{Disorder dependent }\\
\hline
\end{tabular}
\end{center}
}
\end{widetext}

\acknowledgments
This work was supported by funds from Dept. of Science and Technology (Nanomission), Govt. of India, Grant No. SR/NM/NS-1101/2011.
  
\end{document}